# TractoFormer: A Novel Fiber-level Whole Brain Tractography Analysis Framework Using Spectral Embedding and Vision Transformers*

Fan Zhang[1], Tengfei Xue[1,2], Weidong Cai[2], Yogesh Rathi[1], Carl-Fredrik Westin[1], and Lauren J O'Donnell[1]

[1] Harvard Medical School, Boston, USA
[2] The University of Sydney, NSW, Australia

**Abstract.** Diffusion MRI tractography is an advanced imaging technique for quantitative mapping of the brain's structural connectivity. Whole brain tractography (WBT) data contains over hundreds of thousands of individual fiber streamlines (estimated brain connections), and this data is usually parcellated to create compact representations for data analysis applications such as disease classification. In this paper, we propose a novel parcellation-free WBT analysis framework, *TractoFormer*, that leverages tractography information at the level of individual fiber streamlines and provides a natural mechanism for interpretation of results using the attention mechanism of transformers. TractoFormer includes two main contributions. First, we propose a novel and simple 2D image representation of WBT, *TractoEmbedding*, to encode 3D fiber spatial relationships and any feature of interest that can be computed from individual fibers (such as FA or MD). Second, we design a network based on vision transformers (ViTs) that includes: 1) data augmentation to overcome model overfitting on small datasets, 2) identification of discriminative fibers for interpretation of results, and 3) ensemble learning to leverage fiber information from different brain regions. In a synthetic data experiment, TractoFormer successfully identifies discriminative fibers with simulated group differences. In a disease classification experiment comparing several methods, TractoFormer achieves the highest accuracy in classifying schizophrenia vs control. Discriminative fibers are identified in left hemispheric frontal and parietal superficial white matter regions, which have previously been shown to be affected in schizophrenia patients.

## 1 Introduction

Diffusion MRI (dMRI) tractography is an advanced imaging technique that enables *in vivo* reconstruction of the brain's white matter (WM) connections [1]. Tractography provides an important tool for quantitative mapping of the brain's connectivity using measures of connectivity or tissue microstructure [2]. These measures have shown promise as potential biomarkers for disease classification using machine learning [3–5], which can improve our understanding of the brain in health and disease [6].

* This work is supported by the following NIH grants: R01MH119222, R01MH125860, P41EB015902, R01MH074794.

Defining a good data representation of tractography for machine learning is still an open challenge, especially at the fiber level. Performing whole brain tractography (WBT) on one individual subject can generate hundreds of thousands (or even millions) of fiber streamlines. WBT data is usually parcellated to create compact representations for data analysis applications. While most popular analyses of the brain's structural connectivity rely on coarse-scale WM parcellations [2], recent studies have demonstrated the power of analyzing WBT at much finer scales of parcellation using high-resolution connectomes [7,8]. While such approaches enable WBT analysis at a very high resolution (e.g., a 32k×32k connectivity matrix), they are still quite high-dimensional and not able to represent information directly extracted from individual fibers.

Another challenge in machine learning for tractography analysis is the limited sample size (number of subjects) of many dMRI datasets. Developing data augmentation methods to increase sample size is a known challenge in structural connectivity research [9]. Small sample sizes limit the use of recently proposed advanced learning techniques such as Transformers [10] and Vision Transformers (ViTs) [11], which are highly accurate [12] but usually require a large number of samples to avoid overfitting [13].

Finally, an important challenge in deep learning for neuroimaging is to be able to pinpoint location(s) in the brain that are predictive of disease or affected by disease [14]. While interpretability is a well-known challenge in deep learning [15,16], newer methods such as ViTs have shown advances in interpretability for vision tasks [17,18].

In this paper, we propose a novel parcellation-free WBT analysis framework, *TractoFormer*, that leverages tractography information at the level of individual fiber streamlines and provides a natural mechanism for interpretation of results using the self-attention scheme of ViTs. TractoFormer includes two main contributions. First, we propose a novel 2D image representation of WBT, referred to as *TractoEmbedding*, based on a spectral embedding of fibers from tractography. Second, we propose a ViT-based network that performs effective and interpretable group classification. In the rest of this paper, we first describe the TractoFormer framework, then we illustrate its performance in two experiments: classification of synthetic data with true group differences, and disease classification between schizophrenia and control.

# 2 Methods

## 2.1 Diffusion MRI Datasets and Tractography

We use two dMRI datasets. The first dataset is used to create the embedding space and includes data from 100 subjects (29.1±3.7 years; 54 F, 46 M) from the Human Connectome Project (www.humanconnectome.org) [19], with 18 b=0 and 90 b=3000 images, TE/TR=89/5520ms, resolution=1.25×1.25×1.25mm$^3$. The second dataset is used for experimental evaluations and includes data from 103 healthy controls (HCs) (31.1±8.7; 52 F, 51 M) and 47 schizophrenia (SCZ) patients (35.8±8.8; 36 F and 11 M) from the Consortium for Neuropsychiatric Phenomics (CNP) (https://openfmri.org/dataset/ds000030) [20], with 1 b=0 and 64 b=1000 images, TE/TR=93/9000 ms, resolution=2×2×2mm[3]. WBT is performed using the two-tensor unscented Kalman filter (UKF) method [21,22] (via SlicerDMRI [23,24]) to generate about one million fibers per subject. UKF has been successful in neuroscientific applications such as disease classification [25] and population statistical comparison [26], and it allows estimation of fiber-specific microstructural properties (including FA and MD). Fiber tracking parameters are as in [27]. Tractography from the 100 HCP are co-registered, followed by alignment of each CNP WBT using a tractography-based registration [28].

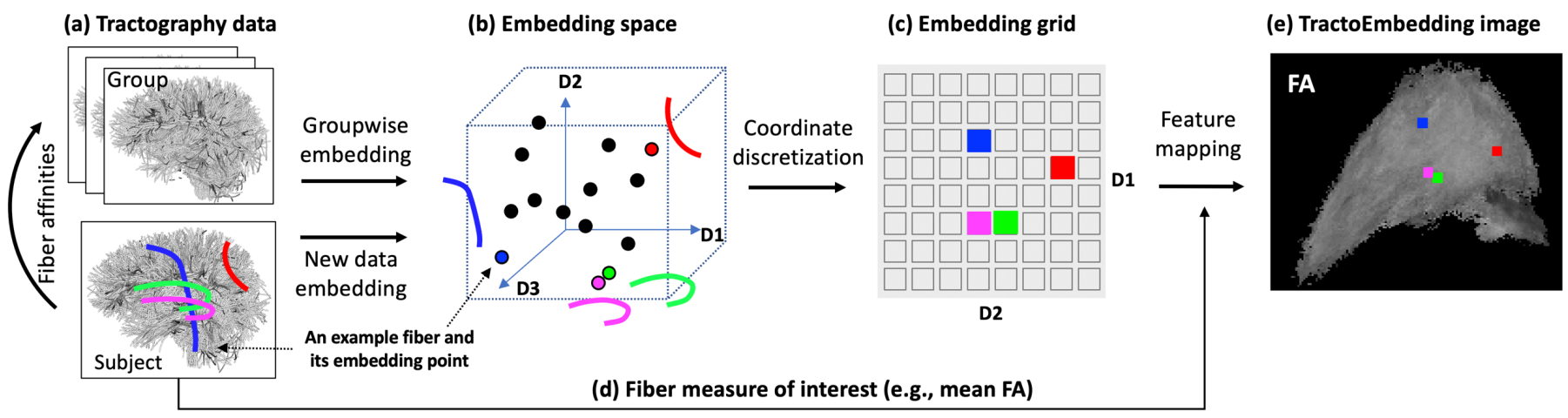

**Fig. 1.** TractoEmbedding overview. Each input fiber in WBT (a) is represented as a point in a latent embedding space (b), where nearby points correspond to spatially proximate fibers. Then, embedding coordinates of all points (fibers) are discretized onto a 2D grid, where points with similar coordinates are mapped to the same or nearby pixels (c). Next, features of interest from each fiber (e.g., mean fiber FA) are mapped (d) as the intensity of the pixel corresponding to that fiber. This generates a 2D image representation, i.e., a TractoEmbedding image (e).

### 2.2 TractoEmbedding: A 2D Image Representation of WBT

The TractoEmbedding process includes three major steps (illustrated in Fig. 1). First, we perform spectral embedding to represent each fiber in WBT as a point in a latent space. Spectral embedding is a learning technique that performs dimensionality reduction based on the relative similarity of each pair of points in a dataset, and it has been successfully used for tractography computing tasks such as fiber segmentation [29], fiber clustering [30], and tract atlas creation [27]. To enable a robust and consistent embedding of WBT data from different subjects for population-wise analysis, we first create a groupwise embedding space using a random sample of fibers from co-registered tractography data from 100 subjects (see Sec. 2.1 for data details). This process uses spectral embedding [31] with a pairwise fiber affinity based on mean closest point distance [30,32]. Next, to embed new WBT data, it is aligned to the 100-subject data [28], followed by computing pairwise fiber affinities to the population tractography sample. Then, each fiber of the new WBT data is spectrally embedded into the embedding space, resulting in an embedding coordinate vector for each fiber. We note that our process of spectral embedding is similar to that used for tractography clustering [30] and we refer the readers to [30] for details.

In the second step, the coordinates of each fiber of the new WBT data are discretized onto a 2D grid for creation of an image. Each dimension of the embedding coordinate vector corresponds to the eigenvectors of the affinity matrix sorted in descending order.

A higher order indicates a higher importance of the coordinate to locate the point in the embedding space. A previous work has applied embedding coordinates for effective visualization of tractography data [33]. In our study, we choose the first two dimensions for each point and discretize them onto a 2D embedding grid[1]. A grid size parameter defines the image resolution.

In the third step, we map the measure of interest associated with each fiber to the corresponding pixel on the embedding grid as its intensity value. This generates a 2D image, i.e., the TractoEmbedding image. When multiple fibers that are spatially proximate are mapped to the same voxel, we can compute summary statistics from these fibers, such as max, min, and mean (mean is used in our experiments).

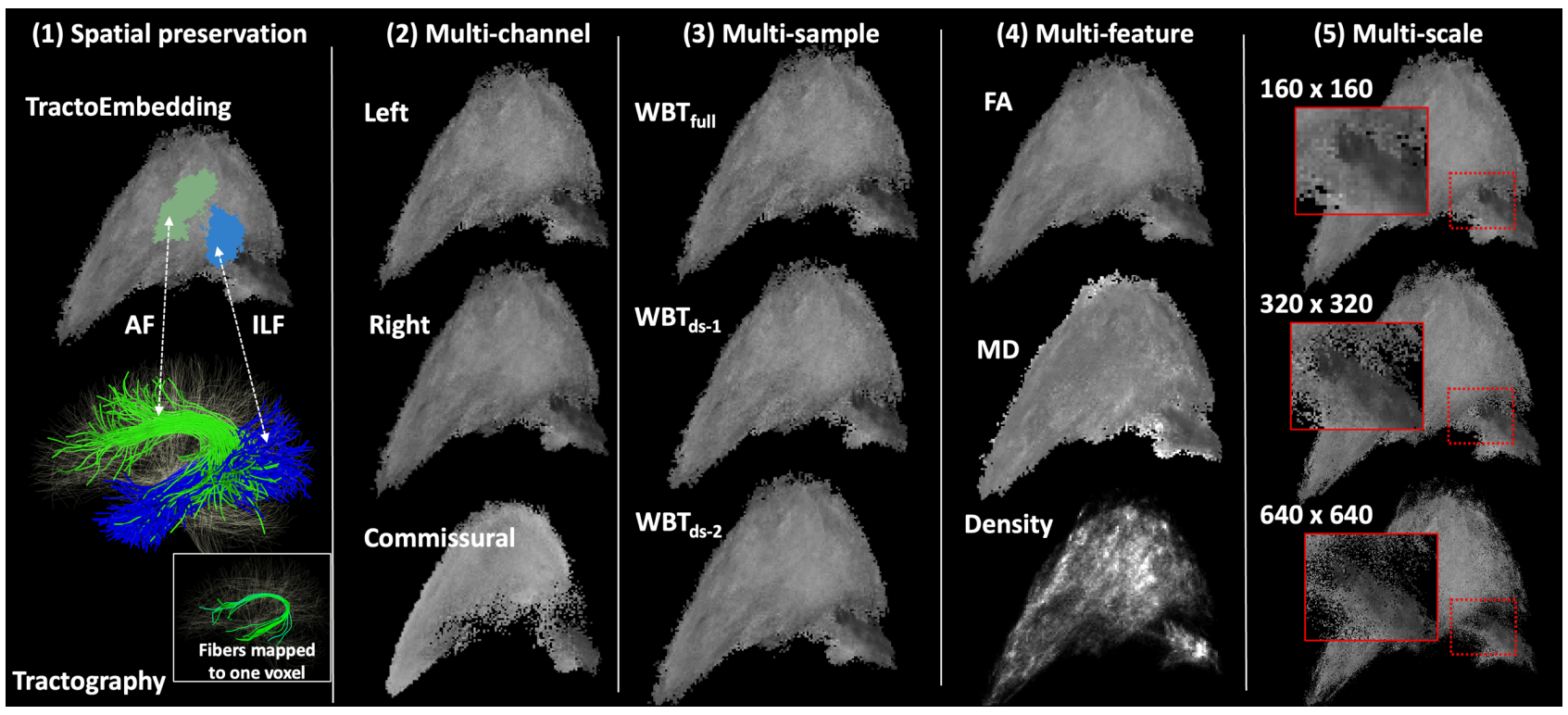

**Fig. 2.** TractoEmbedding images generated from the left hemisphere data of one randomly selected CNP subject. (1) Spatially proximate fibers from the same anatomical tracts are mapped to nearby pixels using TractoEmbedding. (2) Fibers from the left hemisphere, right hemisphere and commissural regions can be used individually to create a multi-channel image. (3) Multiple TractoEmbedding images are generated using the full WBT and two random samples (80% of the full WBT). (4) Multiple TractoEmbedding images are generated using different features of interest, including the mean FA per fiber, the mean MD per fiber, and the number of fibers mapped to each voxel. (5) Multiple TractoEmbedding images are generated at different resolutions (scales). Inset images give a zoomed visualization of a local image region.

TractoEmbedding has several advantages (as illustrated in Fig. 2). *First*, TractoEmbedding is a 2D image that preserves the relative spatial relationship of every fiber pair in WBT in terms of the pixel neighborship in the 2D image (Fig. 2(1)). In this way, TractoEmbedding enables image-based computer vision techniques such as CNNs and ViT to leverage fiber spatial similarity information. (In the case where multiple fibers are mapped to the same voxel, to quantify the similarity of such fibers, we computed the mean pairwise fiber distance (MPFD) across the fibers. The average of MPFDs across all voxels with multiple fibers is 5.7 mm, which is a low value representing highly similar fibers through the same voxel.) *Second*, TractoEmbedding enables a multi-channel representation where each channel represents fibers from certain brain regions. This allows independent and complementary analysis of WBT anatomical regions, such as the left hemispheric, the right hemispheric and the commissural fibers in our current study. Thus, the TractoEmbedding is a 3-channel 2D image (Fig. 2(2)). *Third*, multiple TractoEmbedding images are generated by performing random downsampling of each subject's input WBT (Fig. 2(3)). This naturally and effectively increases training sample size for data augmentation for learning-based methods, which is particularly important for methods that require a large number of samples. *Fourth*, TractoEmbedding can be generally used to encode any possible features of interest that can be computed at the level of individual tractography streamlines (Fig. 2(4)). This enables TractoEmbedding's application in various tractography-based neuroscientific studies where particular WM properties are of interest. *Fifth*, TractoEmbedding allows a WBT representation at different scales in terms of the resolution of the embedding grid (Fig. 2(5)). With a low resolution, multiple fibers tend to be mapped into the same voxel, enabling WBT analysis at a coarse-scale fiber parcel level; with a high resolution, an individual fiber (or a few fibers) is mapped to any particular voxel, enabling WBT analysis at a fine-scale individual fiber level.

[1] While embeddings from the first 3 dimensions can be used to generate 3D TractoEmbedding images, our unpublished results show that this decreases group classification performance potentially due to the data sparsity where many voxels on the 3D grid do not have any mapped fibers.

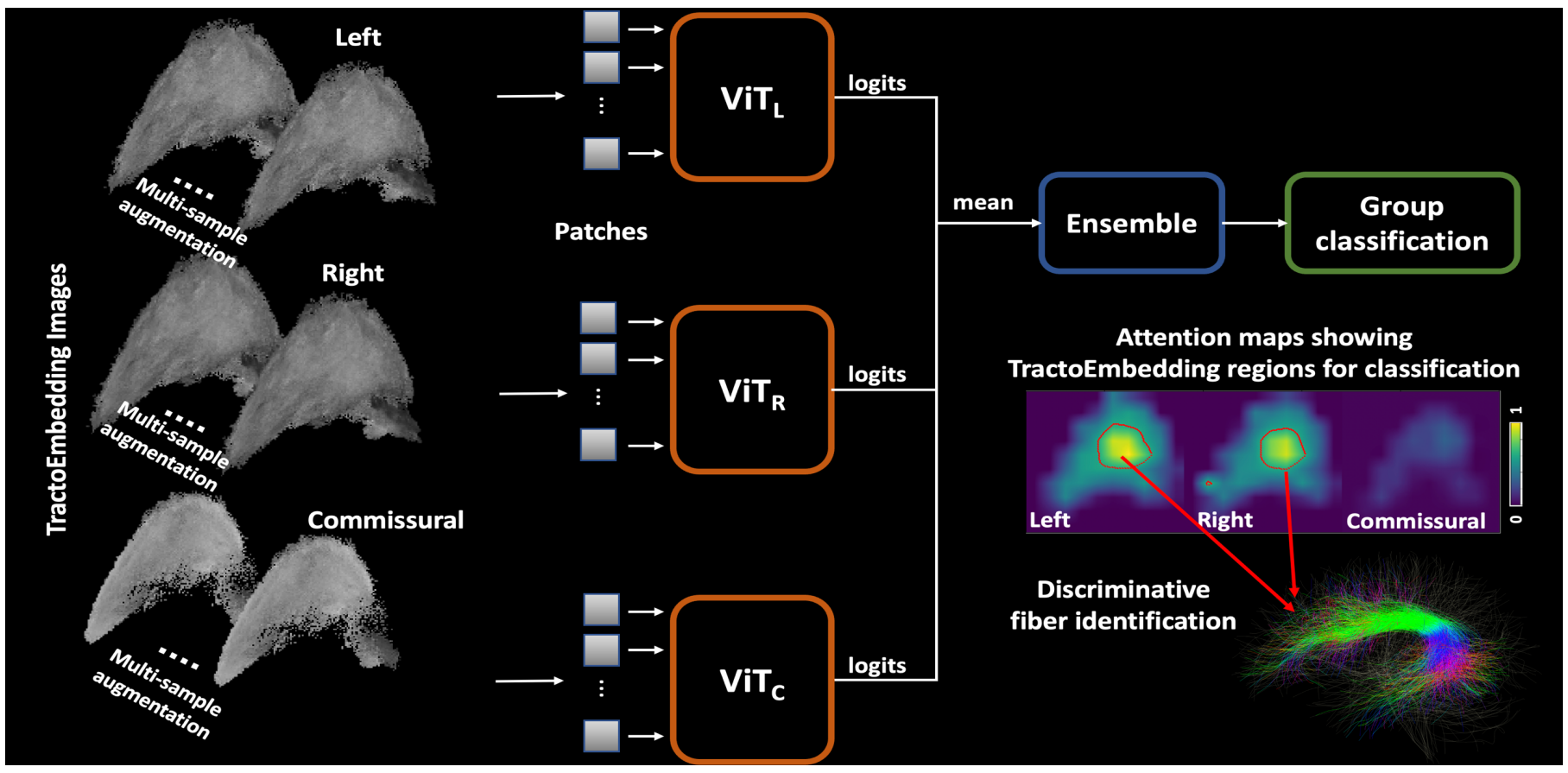

**Fig. 3.** TractoFormer framework including an ensemble ViT network with input multi-channel TractoEmbedding images using multi-sample data augmentation. Attention maps are computed from ViTs for identification of fibers that are discriminative for classification.

## 2.3 TractoFormer: A ViT-based Framework for Group Classification

Fig. 3 shows the proposed TractoFormer architecture, which leverages an ensemble of three ViTs to process the three-channel input TractoEmbedding images. Our design aims to address the aforementioned challenges of sample size/overfitting and interpretability. First, we leverage the multi-sample data augmentation (Fig. 2(3)) to reduce the known overfitting issue of ViTs on small sample size datasets [13]. Second, we leverage the self-attention scheme in ViT to identify discriminative fibers that are most useful to differentiate between groups. The interpretation of the ViT attention maps [11] is aided by our proposed multi-channel architecture, which can enable inspection of the independent contributions of different brain regions.

In detail, for each input channel, we use a light-weight ViT architecture (see Sec. 2.4 for details). An ensemble of the predictions is performed by averaging the logit outputs across the ViTs. For data augmentation, for each input subject, we create 100 TractoEmbedding images using randomly downsampled WBT data (80% of the fibers). For interpretation of results, in each ViT we compute the average of the attention weights for each token across all heads per layer, then recursively multiply the averaged weights for the first to the last layer, and finally map the joint token attention scores back to the input image space[2]. This generates an attention score map where the values indicate the importance of the corresponding pixels when classifying the TractoEmbedding image (as shown in Fig 3). We identify the pixels with higher scores using a threshold *T*, and then identify the fibers that are mapped to these pixels when performing TractoEmbedding. These fibers are thus the ones that are highly important when classifying the TractoEmbedding image. We refer to the identified fibers as *the discriminative fibers*.

### 2.4 Implementation Details and Parameter Setting

Our method is implemented using PyTorch (v1.7.1) [34]. For each ViT, we use 3 layers with 8 heads, a hidden size of 128, and a dropout rate of 0.2 (grid search for {3, 4, 5}, {4,6,8}, {128,256}, and {0.2, 0.3}, respectively). Adam [35] is used for optimization with a learning rate 1e-3 and a batch size 64 for a total of 200 epochs. Early stopping is adopted when there is no accuracy improvement in 20 continuous epochs. 5-fold cross-validation is performed for each experiment below and the mean accuracy and F1 scores are reported. *T* is set to be the mean+2 stds of the scores in an attention map. The computation is performed using NVIDIA GeForce 1080 Ti. On average, each epoch (training and validation) takes ~30 seconds with 2GB GPU memory usage when using data augmentation and 160x160 resolution. The code will be made available upon request.

### 2.5 Experimental Evaluation

**Exp 1: Synthetic data.** The goal is to provide a proof-of-concept evaluation to assess if the proposed TractoFormer can 1) successfully classify groups with true WM differences and 2) identify the fibers with group differences in the WBT data for interpretation. To do so, we create a realistic synthetic dataset with true group differences, as follows. From the 103 CNP HC data, we add white Gaussian noise (signal-to-noise ratio at 1 [36,37]) to the actual measured mean FA value of each fiber in the WBT data. Repeating this process twice generates two synthetic groups of G1 and G2, each with 103 subjects. We then modify the mean FA of the fibers belonging to the corticospinal tract (CST) (a random tract selected for demonstration) in G2 to have a true group difference. To do so, we decrease the mean FA of each CST fiber in G2 by 20%, a synthetic change suggested to introduce a statistically significant difference in tractography-based group comparison analysis [36]. We apply the TractoFormer to this synthetic data to perform group classification and identify the discriminative fibers.

[2] Following instructions from: https://github.com/jeonsworld/ViT-pytorch

**Exp 2: Disease classification between HC and SCZ.** The goal is to evaluate the proposed TractoFormer in a real neuroscientific application for brain disease classification. Previous studies have revealed widespread WM changes in SCZ patients using dMRI techniques [38]. In our paper, we apply TractoFormer to investigate the performance of using tractography data to classify between HC and SCZ in the CNP dataset. For interpretation purposes, we compute a group-wise attention map by averaging the attention maps from all subjects that are classified as SCZ, from which the discriminative TractoEmbedding pixels and discriminative fibers are identified. We compare our method with three baseline methods. The first one performs group classification using fiber parcel level features and a 1D CNN network [39], referred to as the *FC-1DCNN* method. Briefly, for each subject, WBT parcellation is performed using a fiber clustering atlas [27], resulting in a total of 1516 parcels per subject. The mean feature of interest (i.e., FA or MD) along each parcel is computed, leading to a 1D feature vector with 1516 values per subject. Then, a 1D CNN is applied to the feature vectors to perform group classification. For parameters, we follow the suggested settings in the author's implementation[3]. The second method performs group classification using track-density images (TDI) [40] and 3D ResNet [41], referred to as the *TDI-3DResNet* method. Briefly, a 3D TDI, where each voxel represents streamline count, is generated per subject and fed into a 3D ResNet for group classification. The third baseline method performs group classification using TractoEmbedding images, but instead of using the proposed ViT, it applies ResNet [41], a classic CNN architecture that has been shown to be highly successful in many applications. We refer to this method as *ResNet*. For the ResNet and TractoFormer methods, we perform classification with and without data augmentation. We also provide interpretability results using Class Activation Maps (CAMs) [42] in ResNet.

## 3 Results and Discussion

**Exp 1: Synthetic data.** TractoFormer achieved, as expected, 100% group classification accuracy because of the added synthetic feature changes to G2. Fig. 4 shows the identified discriminative fibers in one example G2 subject based on its subject-specific attention map and the G2-group-wise attention map. The discriminative fibers are generally similar to the CST fibers with synthetic changes.

---

[3] https://github.com/H2ydrogen/Connectome_based_prediction

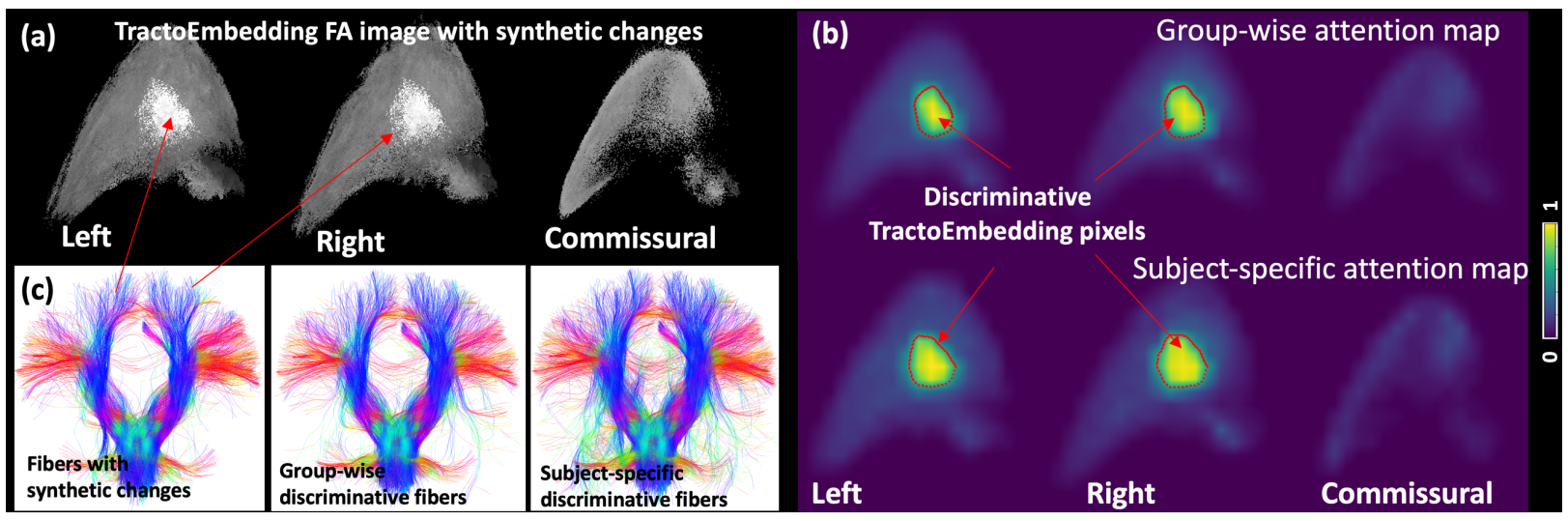

**Fig. 4.** (a) TractoEmbedding FA images of one example G2 subject (320×320). (b) G2-group-wise and subject-specific attention maps (discriminative threshold in red). (c) Identified discriminative fibers, with comparison to the CST fibers with synthetic changes.

**Exp 2: Disease classification between HC and SCZ.** Table 1 shows the classification results of each compared method. In general, the FA measure gives the best result. The FC-1DCNN method generates lower accuracy and F1 scores than the methods that benefit from data augmentation. Regarding the 3 TractoEmbedding-based methods, we can observe that including data augmentation greatly improves the classification performance. The ensemble architecture gives the best overall result (at resolution 160×160 with FA feature), with a mean accuracy of 0.849 and a mean F1 of 0.770. Fig. 5 gives a visualization of the discriminative fibers from group-wise and subject-specific attention maps. In general, our results suggest that the superficial fibers in the frontal and parietal lobes have high importance when classifying SCZ and HC under study. Multiple studies have suggested these white matter regions are affected in SCZ [43–45]. In *ResNet* (at resolution 160×160 with FA feature), CAM identifies the fibers related to the brainstem and cerebellum. The ViT- and ResNet-based methods focus on different brain regions, possibly explaining the accuracy difference of the two methods.

Table 1. Comparison across different methods: the mean accuracy and the mean F1 (the first and second values, respectively, per cell) across the cross-validation are reported.

| Method | Resolution | FA | | MD | | Density | |
|---|---|---|---|---|---|---|---|
| *FC-1DCNN* | n.a. | 0.808/0.669 | | 0.780/0.636 | | 0.767/0.603 | |
| *TDI-3DCNN* | n.a. | n.a. | | n.a. | | 0.764/0.589 | |
| Data Augmentation | | *no aug* | *with aug* | *no aug* | *with aug* | *no aug* | *with aug* |
| *ResNet* | 80×80 | .719/.491 | .819/.751 | .758/.607 | .753/.659 | .630/.506 | .744/.634 |
| | 160×160 | .712/.544 | **.829/.753** | .764/.525 | .769/.662 | .742/.580 | .761/.674 |
| | 320×320 | .653/.524 | .778/.685 | .761/.572 | .703/.604 | .728/.533 | .694/.524 |
| *TractoFormer (stack input)* | 80×80 | .804/.751 | .808/.738 | .649/.606 | .774/.695 | .741/.702 | .774/.701 |
| | 160×160 | .716/.506 | **.816/.732** | .724/.616 | .733/.682 | .708/.486 | .808/.754 |
| | 320×320 | .783/.480 | .783/.653 | .791/.710 | .691/.635 | .783/.480 | .811/.721 |
| *TractoFormer (ensemble)* | 80×80 | .788/.623 | .824/.758 | .716/.600 | .766/.724 | .791/.533 | .808/.743 |
| | 160×160 | .783/.480 | **.849/.770** | .808/.698 | .841/.732 | .733/.525 | .799/.689 |
| | 320×320 | .783/.480 | .758/.645 | .816/.619 | .845/.742 | .741/.489 | .801/.726 |

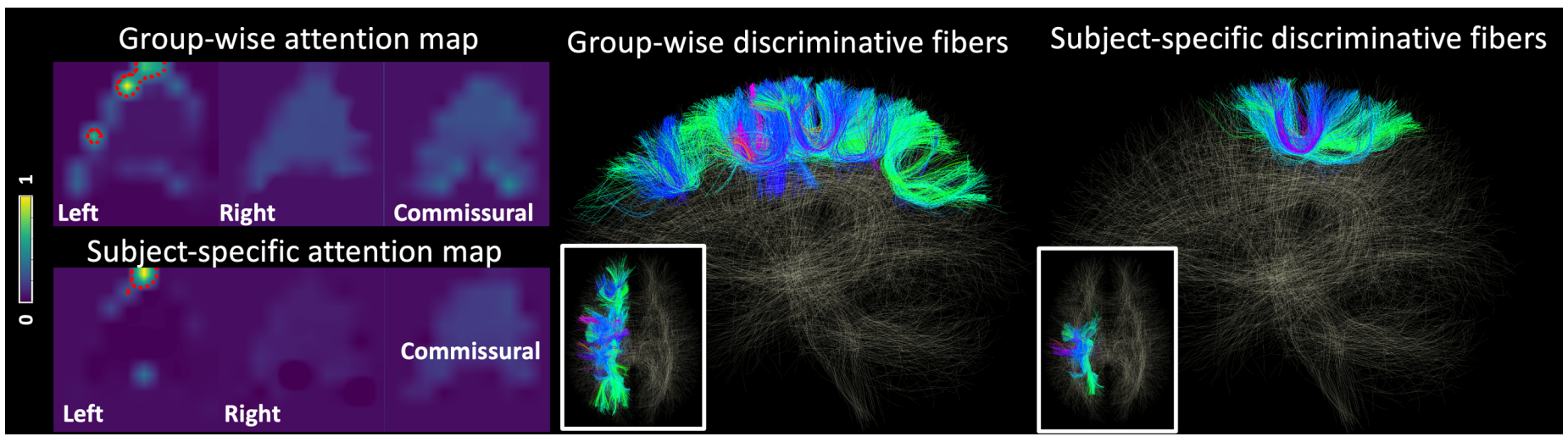

**Fig. 5.** Discriminative fibers identified in the disease classification (SCZ vs HC) experiment, corresponding to the best performing results using FA and resolution 160×160.

# 4 Conclusion

We present a novel parcellation-free WBT analysis framework, *TractoFormer*, which leverages tractography information at the level of individual fiber streamlines and provides a natural mechanism for interpretation of results using attention. We propose random sampling of tractography as an effective data augmentation strategy for small sample size WBT datasets. Future work could include an investigation of ensembles of different fiber features in the same network, multi-scale learning to use TractoEmbedding images with different resolutions together, and/or combination with advanced computer vison data augmentation methods. Overall, TractoFormer suggests the potential for deep learning analysis of WBT represented as images.